\documentclass[letterpaper, 10 pt, conference]{ieeeconf}
\IEEEoverridecommandlockouts
\usepackage[utf8]{inputenc}
\usepackage{tabularx}
\usepackage{float}
\usepackage[caption=false]{subfig}
\usepackage{epsfig}
\usepackage{epstopdf}
\usepackage{bbm}
\usepackage{amsmath} 
\usepackage{amssymb}  
\usepackage{bm}
\usepackage{color}
\DeclareMathOperator*{\argmin}{arg\,min}

\usepackage{algorithm}
\usepackage{algorithmicx}
\usepackage{algpseudocode}

\usepackage{url}

\let\labelindent\relax
\usepackage{enumitem}

\renewcommand{\baselinestretch}{0.9686}

\title{\LARGE \bf
Action Governor for Discrete-Time Linear Systems \\ with Non-Convex Constraints
}

\author{Nan Li$^{1}$, Kyoungseok Han$^{2}$, Anouck Girard$^{1}$, H. Eric Tseng$^{3}$,  Dimitar Filev$^{3}$, Ilya Kolmanovsky$^{1}$
\thanks{*This work is supported by Ford Motor Company.}
\thanks{$^{1}$N. Li, A. Girard, and I. Kolmanovsky are with the Department of Aerospace Engineering, University of Michigan, Ann Arbor, MI 48105, USA. Emails: \{\tt\small{nanli, anouck, ilya}\}@umich.edu}
\thanks{$^{2}$K. Han is with the School of Mechanical Engineering,~Kyungpook National University, Daegu 41566, South Korea. Email: \tt\small{kyoungsh}@knu.ac.kr}
\thanks{$^{3}$H. E. Tseng and D. Filev are with Ford Motor Company, Dearborn, MI 48126, USA. E-mails: \{\tt\small{htseng, dfilev}\}@ford.com}
}

\begin{document}

\maketitle
\thispagestyle{empty}
\pagestyle{empty}

\begin{abstract}
This paper introduces an add-on, supervisory scheme, referred to as {\it Action Governor} (AG), for discrete-time linear systems to enforce exclusion-zone avoidance requirements. It does so by monitoring, and minimally modifying when necessary, the nominal control signal to a constraint-admissible one. The AG operates based on set-theoretic techniques and online optimization. This paper establishes its theoretical foundation, discusses its computational realization, and uses two simulation examples to illustrate its effectiveness.
\end{abstract}

\section{Introduction}\label{sec:1}

Many hard specifications for controlled dynamic systems can be imposed in the form of pointwise-in-time state and control constraints. For instance, such constraints can represent variable bounds to ensure safe and efficient system operation, actuator limits, as well as collision/obstacle avoidance requirements. To address such constraints, one route is to take them into account when (re-)designing the controller, e.g., via correct-by-construction synthesis using controlled invariant sets \cite{nilsson2015correct} or control Lyapunov/barrier functions \cite{tee2009barrier,ames2016control}, or via the model predictive control framework \cite{camacho2013model,borrelli2017predictive}. Another route is to augment a nominal controller with constraint-handling capability via add-on, supervisory schemes \cite{garone2017reference,chen2019enhancing,li2018safe}. This second route may be preferable in many practical circumstances. For instance, a well-designed, legacy controller may already exist while new specifications are imposed to the system. In this case, the second route can preserve many performance characteristics of the legacy controller, such as stability, frequency-domain responses, etc., and thus significantly reduce the tuning complexity compared to re-designing the controller as in the first route~\cite{garone2017reference}.

The Reference Governor (RG) has been shown to be an effective scheme, along the second route, to manage pointwise-in-time state and control constraints \cite{garone2017reference}. The RG monitors and modifies the reference signal, which is typically an input to the nominal controller and defines the control objective, to enforce these constraints. Alternatively, supervision and modification may also be done to the output of the nominal controller, such as in the approaches of \cite{chen2019enhancing,li2018safe}.

In this paper, we propose a novel add-on, supervisory scheme, referred to as {\it Action Governor} (AG), for discrete-time linear systems to enforce pointwise-in-time constraints. Following the ideas of \cite{chen2019enhancing,li2018safe}, the AG enforces constraints by monitoring, and minimally modifying when necessary, the nominal control signal to a constraint-admissible one. 

In particular, we focus on exclusion-zone avoidance requirements, where the zone to avoid is modeled as a convex, polytopic set in the state space. Note that in this case the feasible region is in general non-convex. This differentiates our problem setting from the ones treated in \cite{chen2019enhancing,li2018safe}. The particular consideration for convex exclusion zones is motivated by those obstacle avoidance scenarios frequently encountered in mobile robot path planning problems \cite{latombe2012robot,bouguerra2019viability,hermand2018constrained} as well as vehicle and pedestrian avoidance scenarios in autonomous vehicle control problems \cite{chen2017fast}. 

The AG operation is based on set-theoretic techniques and online optimization. Although there has been a rich literature on set-theoretic methods in control covering theoretical properties, computational aspects and application scenarios, most of previous works treated the case where the feasible region is assumed to be compact and convex \cite{kolmanovsky1998theory,blanchini1999set,rakovic2007optimized,rakovic2010parameterized}. In contrast, the feasible region in our problem setting is in general neither bounded nor convex. 

Although the RG and our proposed AG are both prediction-based supervisory schemes and non-convex constraints have also been considered within the RG framework \cite{tedesco2014collision,lucia2018hybrid,romagnoli2019feedback}, our AG scheme has the following distinguishing features: 1) The AG modifies the output of the nominal controller, whereas the RG modifies the reference input to the controller. 2) Unlike the RG, the AG does not restrict the signal modification over the prediction horizon to a constant. A direct consequence is that the AG yields a larger feasible set, leaving greater flexibility to control, and thereby, can potentially achieve better control performance (as illustrated by an example in Section~\ref{sec:4}). 3) The RG typically assumes the closed-loop system (plant + controller) to be linear + time-invariant (LTI), whereas the AG assumes only the plant to be LTI, i.e., permitting the controller to be nonlinear and evolving with time. This allows controller variability, as well as online learning of the control policy, without needing to redesign/retune the AG.

On the other hand, graph search-based, sampling-based, and potential field-based approaches have been extensively used in path planning for mobile robots with obstacle avoidance requirements \cite{latombe2012robot,elbanhawi2014sampling}. The first two typically use simplified kinematic models for motion prediction and leave the control of system dynamics to a lower level. Although potential field-based approaches can deal with dynamic models, they do not easily lend themselves to theoretical guarantees. In contrast, our AG approach handles dynamic models in the form of discrete-time linear systems and provides theoretical safety guarantees under suitable assumptions.

In summary, the contributions of this paper include: 1)~establishing the theoretical foundation of the AG scheme, 2)~discussing its computational realization, including its offline computational tasks and two online optimization algorithms of different complexity, and 3) illustrating its operation and effectiveness using an automotive and a mobile robot related examples.

\section{Problem Statement}\label{sec:2}

In this paper, we consider a system represented by a discrete-time linear model as follows:
\small
\begin{equation}\label{equ:system_1}
x(k+1) = f\big(x(k),u(k)\big) = A x(k) + B u(k),
\end{equation}
\normalsize
\!\!\! where $x(k) \in \mathbb{R}^n$ represents the system state at the discrete time instant $k \in \mathbb{Z}_{\ge 0}$, and $u(k) \in \mathbb{R}^m$ represents the control input. We assume that a nominal control policy $\phi$ has been defined for the system
\small
\eqref{equ:system_1},
\begin{equation}\label{equ:control_1}
u_{\phi}(k) = \phi\big(x(k),r(k),k\big),
\end{equation}
\normalsize
\!\!\! where $r(k) \in \mathbb{R}^p$ represents a reference signal determining the control objective. The control policy $\phi$ may be nonlinear and time-varying, which may be due to specific control objectives such as state/control constraints, finite-time convergence requirements, etc.

Furthermore, we assume that the system is subject to an {\it exclusion-zone avoidance} requirement of the form
\small
\begin{equation}\label{equ:constraint_1}
x(k) \notin X_0, \quad \forall\, k \in \mathbb{Z}_{\ge 0}.
\end{equation}
\normalsize
\!\!\! In particular, we assume $X_0$ to be a convex, open and polytopic set,
\small
\begin{equation}\label{equ:constraint_2}
X_0 = \big\{x \in \mathbb{R}^n: G x < g \big\},
\end{equation}
\normalsize
\!\!\! where $G \in \mathbb{R}^{n_g \times n}$ and $g \in \mathbb{R}^{n_g}$. Note that the requirement \eqref{equ:constraint_1} can also be written as $x(k) \in \mathbb{R}^n \setminus X_0$, which, in the case of $X_0$ being convex, is in general non-convex.

The exclusion-zone avoidance requirement \eqref{equ:constraint_1} may not have been incorporated when defining the nominal control policy \eqref{equ:control_1}. The objective of this paper is to develop a control algorithm to enforce \eqref{equ:constraint_1}.

\section{Action Governor}\label{sec:3}

The solution that we propose is a supervisory scheme, referred to as {\it Action Governor} (AG), which monitors and minimally modifies, if necessary, the nominal control signal $u_{\phi}(k)$ to enforce the exclusion-zone avoidance requirement~\eqref{equ:constraint_1}.

In particular, the AG operates based on the following constrained optimization problem:
\small
\begin{subequations}\label{equ:AG_1}
\begin{align}
    u(k) =&\, \argmin_{u \in U} \|u - u_{\phi}(k)\|_S^2, \\
    &\text{ subject to } Ax(k) + Bu \in X_{\text{safe}},
\end{align}
\end{subequations}
\normalsize
\!\!\!\! where the set $U \subset \mathbb{R}^m$ represents the range of control authority, the function $\|\cdot\|_S = \sqrt{(\cdot)^{\top} S (\cdot)}$ with $S \in \mathbb{R}^{m \times m}$ being positive-definite penalizes the difference between the modified control signal $u(k)$ and the nominal control signal $u_{\phi}(k)$, and $X_{\text{safe}} \subset \mathbb{R}^n$ is a {\it safe set} which will be introduced later. In particular, we assume $U$ to be a convex, closed and polytopic set as follows, which is a common assumption in the control literature \cite{camacho2013model,borrelli2017predictive,garone2017reference},
\small
\begin{equation}\label{equ:constraint_3}
U = \big\{u \in \mathbb{R}^m: H u \le h \big\},
\end{equation}
\normalsize
\!\!\! where $H \in \mathbb{R}^{n_h \times m}$ and $h \in \mathbb{R}^{n_h}$.

\subsection{Safe Set and Unrecoverable Sets}

To enforce both present and future safety, the safe set $X_{\text{safe}}$ is characterized by the following requirements: For any $x(k) \in X_{\text{safe}}$, there exists $u(k) \in U$ such that
\begin{enumerate}[leftmargin=0cm,itemindent=.5cm,labelwidth=\itemindent,labelsep=0cm,align=left]
\item The next state satisfies the exclusion-zone avoidance requirement \eqref{equ:constraint_1}, i.e., $x(k+1) = Ax(k) + Bu(k) \notin X_0$.
\item Future exclusion-zone avoidance is possible, i.e., given $x(k+1) = Ax(k)+Bu(k)$, there exists a control sequence $\{u(k+1),u(k+2),\dots\} \subset U$ such that $\{x(k+2),x(k+3),\dots\} \cap X_0 = \emptyset$.
\end{enumerate}

The explicit determination of $X_{\text{safe}}$ relies on the following sequence of sets, referred to as {\it unrecoverable sets}, defined sequentially as
\small
\begin{align}\label{equ:constraint_4}
X_k &= X_0 \cup \big\{x \in \mathbb{R}^n:\, \text{for each $u \in U$,}\, Ax + Bu \in X_j \nonumber \\
&\quad\quad\quad\quad \text{ for some $j = 0,\dots,k-1$} \big\} \nonumber \\[-3pt]
=&\, X_0 \cup \Big\{x \in \mathbb{R}^n: Ax + Bu \in \text{$\bigcup_{j = 0}^{k-1}$} X_j,\, \forall\, u \in U \Big\} \\[-3pt]
=&\, X_0 \cup \Big\{x \in \mathbb{R}^n: Ax \in \Big(\bigcup_{j = 0}^{k-1} X_j\Big) \sim BU \Big\}, \quad k = 1,2,\dots \nonumber 
\end{align}
\normalsize
\!\!\! where $\sim$ denotes the P(ontryagin)-difference operation between sets \cite{kolmanovsky1998theory}.

The sets $X_k$ satisfy the following properties, which also explain why they are called ``unrecoverable sets'':

{\em Proposition 1:} If $x(0) \in X_k$, then for any sequence $\{u(0),\cdots,u(k-1)\} \in U \times \cdots \times U$ there exists $0 \le j \le k$ such that $x(j) \in X_0$.

{\em Proof:} The proof is by induction. For $k = 1$, $x(0) \in X_1$ implies either $x(0) \in X_0$ or $x(1) = Ax(0) + Bu(0) \in X_0$ for any $u(0) \in U$. Suppose the statement has been proven for $X_j$, $1 \le j \le k$. For $k+1$, if $x(0) \in X_{k+1}$, then either $x(0) \in X_0$ or for each $u(0) \in U$, $x(1) = A x(0) + B u(0) \in X_j$ for some $0 \le j \le k$, which is by the definition of $X_{k+1}$. For the latter, since $x(1) \in X_j$, by our induction hypothesis, for any $\{u(1), \cdots, u(j)\} \in U \times \cdots \times U$, there exists $0 \le i \le j$ such that $x(i+1) \in X_0$. Thus, we have shown that for any $\{u(0), u(1), \cdots, u(k)\} \in U \times U \times \cdots \times U$, there exists $0 \le j' = i+1 \le j+1 \le k+1$ such that $x(j') \in X_0$. This proves the statement for $k+1$. $\blacksquare$

{\em Proposition 2:} Let $x(0)$ be given. If for any sequence $\{u(0),\cdots,u(k-1)\} \in U \times \cdots \times U$ there exists $0 \le j \le k$ such that $x(j) \in X_0$, then $x(0) \in X_k$.

{\em Proof:} The proof is by induction. For $k = 1$, $x(0) \in X_0$ or $x(1) = Ax(0) + Bu(0) \in X_0$ for any $u(0) \in U$ implies $x(0) \in X_1$, which is by the definition of $X_1$. Suppose the statement has been proven for $k$. For $k+1$, if $x(0) \in X_0$, then $x(0) \in X_{k+1}$, which follows from the fact that $X_0 \subset X_{k+1}$. Otherwise, fix an arbitrary $u(0) \in U$ and let $x(1) = Ax(0) + Bu(0)$. The proposition statement assumes that for any $\{u(1),\cdots,u(k)\} \in U \times \cdots \times U$, there exists $0 \le j \le k$ such that $x(j+1) \in X_0$, which implies $x(1) \in X_k$ by our induction hypothesis. Thus, we have shown that for each $u(0) \in U$, $Ax(0) + Bu(0) \in X_k$, which implies $x(0) \in X_{k+1}$ by the definition of $X_{k+1}$. This proves the statement for $k+1$. $\blacksquare$

Propositions~1 and 2 mean that there exists no solution for the control sequence $\{u(0),\cdots,u(k-1)\} \in U \times \cdots \times U$ to avoid the state trajectory entering the exclusion zone $X_0$ over the steps $0,\dots,k$ if and only if $x(0) \in X_k$. Or equivalently, there exists a control sequence $\{u(0),\cdots,u(k-1)\} \in U \times \cdots \times U$ to avoid $X_0$ over the steps $0,\dots,k$ if and only if $x(0) \in \mathbb{R}^n \setminus X_k$. Furthermore, the following convergence property of $X_k$ holds:

{\em Proposition 3:} For each $k = 0,1,\dots$, $X_k \subset X_{k+1}$, i.e., $X_k$ is an increasing sequence of sets. In turn, $X_{\infty} = \lim_{k \to \infty} X_k$ exists (in the set-theoretic sense) and satisfies $X_k \subset X_{\infty}$ for all $k$.

{\em Proof:} The proof follows from
\small
\begin{align}
    & X_k = X_0 \cup \Big\{x \in \mathbb{R}^n: Ax \in \Big(\bigcup_{j = 0}^{k-1} X_j\Big) \sim BU \Big\} \\[-3pt]
    &\subset X_0 \cup \Big\{x \in \mathbb{R}^n: Ax \in \Big(\bigcup_{j = 0}^{k} X_j\Big) \sim BU \Big\} = X_{k+1}. \quad \blacksquare \nonumber 
\end{align}
\normalsize

Using the unrecoverable sets $X_k$, we define the safe set as $X_{\text{safe}} = \mathbb{R}^n \setminus X_{\infty} = \lim_{k \to \infty} (\mathbb{R}^n \setminus X_k)$. On the basis of Propositions~1 and 2, $X_{\text{safe}}$ has the following properties:

{\em Proposition 4:} For any $x \in X_{\text{safe}}$, it holds that (i) $x \notin X_0$, and (ii) there exists $u \in U$ such that $Ax + Bu \in X_{\text{safe}}$.

{\em Proof:} The former $x \in X_{\text{safe}} \implies x \notin X_0$ follows from
\small
\begin{equation}
X_{\text{safe}} = \mathbb{R}^n \setminus X_{\infty} = \mathbb{R}^n \setminus \big(\bigcup_{k=0}^{\infty} X_k\big) \subset \mathbb{R}^n \setminus X_0.
\end{equation}
\normalsize

For the latter, let $x$ be given and assume that for any $u \in U$, $Ax + Bu \in \mathbb{R}^n \setminus X_{\text{safe}} =  \bigcup_{k=0}^{\infty} X_k$, i.e., $Ax \in \big(\bigcup_{k=0}^{\infty} X_k\big) \sim BU = \bigcup_{k=0}^{\infty} \big(X_k \sim BU \big)$, where the last equality holds according to Lemma~1 in Appendix. This means there must exist $k'$ such that $Ax \in X_{k'} \sim BU = \big(\bigcup_{k=0}^{k'} X_k\big) \sim BU$. Then, according to \eqref{equ:constraint_4}, $x \in X_{k'+1} \subset \bigcup_{k=0}^{\infty} X_k = \mathbb{R}^n \setminus X_{\text{safe}}$. Thus, if $x \in X_{\text{safe}}$, then there must exist $u \in U$ such that $Ax + Bu \notin \mathbb{R}^n \setminus X_{\text{safe}}$. $\blacksquare$

Proposition~4 ensures that if the AG operates based on \eqref{equ:AG_1} to modify the control input $u(k)$, then 1) a feasible solution exists to \eqref{equ:AG_1} for all $k$, and 2) the exclusion-zone avoidance requirement \eqref{equ:constraint_1} is satisfied for all $k$.

Note that the exact determination of $X_{\text{safe}}$ relies on the set $X_k$ iteratively computed according to \eqref{equ:constraint_4} with $k \to \infty$. In practice, $X_{\text{safe}}$ can be approximated by $\tilde{X}_{\text{safe},k'} = \mathbb{R}^n \setminus X_{k'}$ with $k'$ being sufficiently large. Moreover, the following result says that, under a few additional assumptions, such a finitely determinable approximation of $X_{\text{safe}}$ suffices for implementation. 

{\em Proposition 5:} Assume $0 \in U$ and define $R = \bigoplus_{k=0}^{\infty}A^k BU$, where $\oplus$ denotes the Minkowski sum operation of sets \cite{kolmanovsky1998theory}. Suppose 1) there exists $k'$ such that
\small
\begin{equation}\label{equ:AG_20}
R \cap \big( X_{\infty} \setminus X_{k'} \big) = \emptyset,
\end{equation}
\normalsize
\!\!\! 2) $x(0) \in R \cap \tilde{X}_{\text{safe},k'}$, and 3) the AG operates based on
\small
\begin{subequations}\label{equ:AG_2}
\begin{align}
    u(k) =&\, \argmin_{u \in U} \|u - u_{\phi}(k)\|_S^2, \\
    &\text{ subject to } Ax(k) + Bu \in \tilde{X}_{\text{safe},k'}.
\end{align}
\end{subequations}
\normalsize
\!\!\! Then, (i) $x(k) \notin X_0$, and (ii) there exists $u \in U$ such that $Ax(k) + Bu \in \tilde{X}_{\text{safe},k'}$, for all $k = 0,1,\dots$

{\em Proof:} Firstly, (i) follows from $x(k) \in \mathbb{R}^n \setminus X_{k'} \subset \mathbb{R}^n \setminus X_{0}$. Now assume $x(k-1) \in R \cap \tilde{X}_{\text{safe},k'} = R \cap (\mathbb{R}^n \setminus X_{k'})$. Note that \eqref{equ:AG_20} implies
\small
\begin{align}\label{equ:AG_21}
& R \cap \big( X_{\infty} \setminus X_{k'} \big) = R \cap (\mathbb{R}^n \setminus X_{k'}) \cap X_{\infty} = \emptyset \nonumber \\
&\implies R \cap (\mathbb{R}^n \setminus X_{k'}) \subset \mathbb{R}^n \setminus X_{\infty} = X_{\text{safe}}.
\end{align}
\normalsize
\!\!\! Thus, we have $x(k-1) \in X_{\text{safe}}$, which by Proposition~4 ensures the existence of $u \in U$ such that $Ax(k-1) + Bu \in X_{\text{safe}} = \mathbb{R}^n \setminus X_{\infty} \subset \mathbb{R}^n \setminus X_{k'} = \tilde{X}_{\text{safe},k'}$. This proves (ii) for $k-1$. Also, by the invariance of $R$ \cite{kolmanovsky1998theory}, $x(k-1) \in R$ implies $Ax(k-1) + Bu \in R$ for any $u \in U$. Therefore, if the AG operates based on \eqref{equ:AG_2} at $k-1$, we must have $x(k) = Ax(k-1) + Bu(k-1) \in R \cap \tilde{X}_{\text{safe},k'}$. Then, the proof of (ii) for all $k = 0,1,\dots$ is completed by an induction argument. $\blacksquare$

We remark that our definition of the safe set $X_{\text{safe}}$ is similar to the {\it viability kernel} considered in \cite{bouguerra2019viability}. In \cite{bouguerra2019viability}, the computation and utilization of the viability kernel are based on finite state and control spaces (or based on finite discretization of the original spaces). In contrast, we deal with continuous state and control spaces in this paper, and the computational algorithms introduced in what follows do not rely on discretization of the spaces. Furthermore, the theoretical results of this paper, Propositions~1-6, do not appear in \cite{bouguerra2019viability}.

\subsection{Offline and Online Computations}

Given $X_0$ as a convex, polytopic set \eqref{equ:constraint_2}, the sets $X_k$ for $k = 1,2,\dots$ are iteratively computed offline based on the following proposition:

{\em Proposition 6:} Assume $A$ is invertible (see Remark~1) and $U$ is a polytopic set. Then, for each $k = 1,2,\dots$, we have (i) $X_k$ can be represented as the union of a finite number of polytopic sets, i.e., $X_k = \bigcup_{j = 1}^{r_k} X_{k,j}$ where $X_{k,j}$ is a polytopic set for each $j = 1,\dots,r_k$; and (ii) $X_k$ can be numerically computed using Algorithm~1.

{\em Proof:} Assume that $X_{k-1}$ can be represented as the union of a finite number of polytopic sets. Using the fact that $X_j \subset X_{j+1}$ for all $j = 0,1,\dots,k-2$, we can rewrite \eqref{equ:constraint_4} as
\small
\begin{equation}
X_k = X_0 \cup \big\{x \in \mathbb{R}^n: Ax \in X_{k-1} \sim BU \big\}.
\end{equation}
\normalsize
\!\!\! Note that $BU$, as the image of a polytopic set $U$ under the linear transformation $B$, is also a polytopic set. Then, $X_{k-1} \sim BU$ is the P-difference between a finite union of polytopic sets and a polytopic set. Thus, according to Theorem~4.4 of \cite{borrelli2017predictive}, the lines~1-5 of Algorithm~1 compute $\mathcal{G} = X_{k-1} \sim BU$, which is also a finite union of polytopic sets. Then, since $A$ is invertible, the preimage of $\mathcal{G}$ under $A$, $A^{-1} \mathcal{G}$, is still a finite union of polytopic sets. Furthermore, we have
\small
\begin{equation}
A^{-1} \mathcal{G} = \big\{x \in \mathbb{R}^n: Ax \in X_{k-1} \sim BU \big\}.
\end{equation}
\normalsize
\!\!\! Finally, since $X_0$ is a polytopic set, we obtain $X_k = X_0 \cup A^{-1} \mathcal{G}$ is a finite union of polytopic sets. Then, (i) and (ii) are simultaneously proved for $k$. The proof is extended to all $k = 1,2,\dots$ by an induction argument. $\blacksquare$

{\em Remark 1:} We remark that the assumption of $A$ being invertible typically holds true for a practical system, e.g., when the model \eqref{equ:system_1} is discretized from continuous-time dynamics. We also remark that the set operations involved in Algorithm~1, including P-difference between polytopic sets in line~2, and convex hull, set difference, Minkowski sum and union of finite unions of polytopic sets in lines~1,~3-6 can be efficiently computed using corresponding functions of the Multi-Parametric Toolbox 3 (MPT3) \cite{MPT3}.

\begin{algorithm}
        \caption{Offline computation for $X_k$}
        \label{Algorithm1}
        \begin{algorithmic}[1] 
            \Require $A,B,X_0,X_{k-1},U$ 
            \Ensure $X_k$ \\
            $\mathcal{H} \leftarrow \text{convhull}(X_{k-1})$ \\
            $\mathcal{D} \leftarrow \mathcal{H} \sim (BU)$ \\
            $\mathcal{E} \leftarrow \mathcal{H} \setminus X_{k-1}$ \\
            $\mathcal{F} \leftarrow \mathcal{E} \oplus (-BU)$ \\
            $\mathcal{G} \leftarrow \mathcal{D} \setminus \mathcal{F}$ \\
            $X_k \leftarrow X_0 \cup A^{-1} \mathcal{G}$
        \end{algorithmic}
\end{algorithm}

From now on, we assume the AG operates based on $\tilde{X}_{\text{safe},k'} = \mathbb{R}^n \setminus X_{k'}$ for some $k'$. According to Proposition~6, $X_{k'}$ is a finite union of polytopic sets, i.e., can be written as
\small
\begin{equation}
 X_{k'} = \bigcup_{j = 1}^{r_{k'}} \bigcap_{i = 1}^{s_{j}} \big\{x \in \mathbb{R}^n: G_{i,j} x < g_{i,j} \big\},  
\end{equation}
\normalsize
\!\!\! where $G_{i,j} \in \mathbb{R}^{1 \times n}$, $g_{i,j} \in \mathbb{R}$, and in turn,
\small
\begin{equation}
\tilde{X}_{\text{safe},k'} = \mathbb{R}^n \setminus X_{k'} = \bigcap_{j = 1}^{r_{k'}} \bigcup_{i = 1}^{s_{j}} \big\{x \in \mathbb{R}^n: G_{i,j} x \ge g_{i,j} \big\}.  
\end{equation}
\normalsize

Then, the constraint $Ax(k) + Bu \in \tilde{X}_{\text{safe},k'}$ is equivalent to the following set of constraints:
\small
\begin{subequations}\label{equ:AG_MIQP} 
\begin{align}
   & G_{i,j} (Ax(k) + Bu) \ge g_{i,j} - M (1-\delta_{i,j}), \\[9pt]
   & \delta_{i,j} \in \{0,1\},\quad \forall\, i = 1,\dots,s_j,\, \forall\, j = 1,\dots, r_{k'}, \\
   & \sum_{i = 1}^{s_j} \delta_{i,j} = 1,\quad \forall\, j = 1,\dots, r_{k'},
\end{align}
\end{subequations}
\normalsize
\!\!\! where $M>0$ is a sufficiently large positive number.

Thus, the AG online problem \eqref{equ:AG_2} can be solved as a Mixed-Integer Quadratic Programming (MIQP) problem with $(u,\delta_{i,j})$ as the decision variables.

Furthermore, under the following practical assumption, a computationally lighter approach, presented as Algorithm~2, can be used to approximately solve \eqref{equ:AG_2}.

{\em Assumption~1:} A safe-mode control policy $\psi$ has been defined for the system
\eqref{equ:system_1},
\small
\begin{equation}\label{equ:control_21}
u_{\psi}(k) = \psi\big(x(k),k\big),
\end{equation}
\normalsize
\!\!\! possibly being conservative, such that for any $x(k) \in \tilde{X}_{\text{safe},k'}$ we have
\small
\begin{equation}\label{equ:control_22}
Ax(k) + Bu_{\psi}(k) \in \tilde{X}_{\text{safe},k'}.
\end{equation}
\normalsize

Assumption~1 is reasonable for many practical systems. For instance, for an Adaptive Cruise Control (ACC) system where $x(k) \in X_0$ represents a rear-end collision to the preceding vehicle, the safe-mode policy $\psi$ may correspond to an Automatic Emergency Braking (AEB) system.

Algorithm~2 aims to find a feasible point $u(k)$ along the line segment connecting $u_{\phi}(k)$ and $u_{\psi}(k)$ that is as close to $u_{\phi}(k)$ as possible through a bisection method. It is similar to the Algorithm~1 in \cite{cotorruelo2019output}. Two important properties of Algorithm~2 are: 1) algorithm convergence is guaranteed, i.e., $\underline{\lambda}$ and $\overline{\lambda}$ are converging to the same value $\in [0,1]$; and 2) algorithm output $u(k)$ is always feasible, in terms of satisfying that $Ax(k)+Bu(k) \in \tilde{X}_{\text{safe},k'}$. For more details regarding these two properties, the reader is referred to \cite{cotorruelo2019output,bemporad1998reference}. We also remark that the set containment condition in line~4 can be efficiently checked using the {\it PolyUnion.contains()} function of MPT3 \cite{MPT3}.

\begin{algorithm}
        \caption{Online computation for $u(k)$}
        \label{Algorithm1}
        \begin{algorithmic}[1] 
            \Require $A,B,\tilde{X}_{\text{safe},k'},x(k), u_{\phi}(k),u_{\psi}(k)$ 
            \Ensure $u(k)$ \\
            $\underline{\lambda} \leftarrow 0$, $\overline{\lambda} \leftarrow 1$, $\lambda \leftarrow 1$
            \While{$\overline{\lambda} - \underline{\lambda} > \delta$} \\
            \quad\, $u \leftarrow \lambda u_{\phi}(k) + (1-\lambda) u_{\psi}(k)$
            \If{$Ax(k)+Bu \in \tilde{X}_{\text{safe},k'}$} \\
            \quad\quad\,\, $\underline{\lambda} \leftarrow \lambda$ 
            \Else \\
            \quad\quad\,\, $\overline{\lambda} \leftarrow \lambda$
            \EndIf \\
            \quad\, $\lambda \leftarrow (\underline{\lambda} + \overline{\lambda})/2$
            \EndWhile \\
            $u(k) \leftarrow \underline{\lambda} u_{\phi}(k) + (1-\underline{\lambda}) u_{\psi}(k)$
        \end{algorithmic}
\end{algorithm}

\section{Examples}\label{sec:4}

\subsection{Adaptive Cruise Control}

The first example we consider represents an Adaptive Cruise Control (ACC) system for automated highway driving. The relative motion between the leading vehicle and the following ego vehicle is written in discrete-time as
\small
\begin{align}
    \begin{bmatrix} \Delta s(k+1) \\ \Delta v(k+1) \end{bmatrix} = \begin{bmatrix} 1 & \Delta t \\ 0 & 1 \end{bmatrix} \begin{bmatrix} \Delta s(k) \\ \Delta v(k) \end{bmatrix} - \begin{bmatrix} \frac{1}{2} \Delta t^2 \\ \Delta t \end{bmatrix} u(k),
\end{align}
\normalsize
\!\!\! where $\Delta t = 0.25$[s] is the sampling period, $\Delta s$ and $\Delta v$ denote, respectively, the longitudinal distance and relative speed between the leading and the ego vehicles, and $u$ is the control input representing the ego vehicle's acceleration. The following feedback policy is defined as the nominal control,
\small
\begin{equation}\label{equ:acc_u}
    u_{\phi}(k) = K \begin{bmatrix} \Delta s(k) - \Delta s_r(k) \\ \Delta v(k) \end{bmatrix},
\end{equation}
\normalsize
\!\!\! where $\Delta s_r$ is the reference signal and represents the desired car-following distance, and the feedback gain $K$ is computed as the Linear Quadratic Regulator (LQR) gain with $Q = \text{diag}(10,1)$ and $R = 20$.

To avoid rear-end collision and promote passenger comfort, the following constraints are imposed,
\small
\begin{equation}\label{equ:acc_c}
    \Delta s(k) \ge 2\text{[m]}, \quad -2\text{[m/s$^2$]} \le u(k) \le 2\text{[m/s$^2$]}, \quad \forall\, k.
\end{equation}
\normalsize

Clearly, the nominal policy \eqref{equ:acc_u} does not account for the constraints \eqref{equ:acc_c}. We now consider both the applications of AG and RG to enforcing \eqref{equ:acc_c} and compare them.

On the one hand, the constraints \eqref{equ:acc_c} can be handled by AG with considering the following definition for the exclusion set $X_0$,
\small
\begin{equation}
    X_0 = \left\{ \begin{bmatrix} \Delta s \\ \Delta v \end{bmatrix} \in \mathbb{R}^2: \begin{bmatrix} 1 & 0 \\ -1 & 0 \\ 0 & 1 \\ 0 & -1 \end{bmatrix} \begin{bmatrix} \Delta s \\ \Delta v \end{bmatrix} < \begin{bmatrix} 2 \\ M \\ M \\ M \end{bmatrix} \right\},
\end{equation}
\normalsize
\!\!\! where $M>0$ is a sufficiently large positive number, and the control authority set $U = [-2,2]$. The reason for including the 2nd-4th inequalities as ``virtual constraints'' is to make $X_0$ a polytopic set so that enable the numerical computations of Algorithm~1. In this example, the online determination of the control input $u$ is through solving the MIQP \eqref{equ:AG_2} and~\eqref{equ:AG_MIQP}.

On the other hand, the RG considers the closed-loop system after the nominal control \eqref{equ:acc_u} is applied, i.e.,
\small
\begin{align}
\begin{bmatrix} \Delta s(k+1) \\ \Delta v(k+1) \end{bmatrix} &= \left(\begin{bmatrix} 1 & \Delta t \\ 0 & 1 \end{bmatrix} - \begin{bmatrix} \frac{1}{2} \Delta t^2 \\ \Delta t \end{bmatrix} K\right) \begin{bmatrix} \Delta s(k) \\ \Delta v(k) \end{bmatrix} \nonumber \\
&\quad + \begin{bmatrix} \frac{1}{2} \Delta t^2 \\ \Delta t \end{bmatrix} K(1)\, \Delta s_r(k),
\end{align}
\normalsize
\!\!\! where $K(1)$ denotes the first entry of $K$. The RG monitors and modifies, if necessary, the reference signal $\Delta s_r(k)$ to a constraint-admissible one $\Delta s_v(k)$ to enforce constraints. For this, it utilizes a set typically called $O_{\infty}$ \cite{garone2017reference}. In particular, to handle the constraints \eqref{equ:acc_c} and also enable the numerical computation of $O_{\infty}$, we consider the following constraints on the state-reference pair,
\small
\begin{equation}
    \begin{bmatrix} -1 & 0 \\ K(1) & K(2) \\ -K(1) & -K(2) \\ 1 & 0 \\ 0 & 1 \\ 0 & -1 \end{bmatrix} \begin{bmatrix} \Delta s(k) \\ \Delta v(k) \end{bmatrix} + \begin{bmatrix} 0 \\ -K(1) \\ K(1) \\ 0 \\ 0 \\ 0 \end{bmatrix} \Delta s_v(k) \le \begin{bmatrix} -2 \\ 2 \\ 2 \\ M \\ M \\ M \end{bmatrix}.
\end{equation}
\normalsize

We consider the following initial condition and constant reference signal,
\small
\begin{equation*}
    \big(\Delta s(0), \Delta v(0)\big) = \big(18\text{[m]}, -4\text{[m/s]}\big), \quad \Delta s_r(k) \equiv 2.5\text{[m]}.
\end{equation*}
\normalsize

\begin{figure}[h!]
\begin{center}
\begin{picture}(245.0, 112.0)
\put(  -5,  -10){\epsfig{file=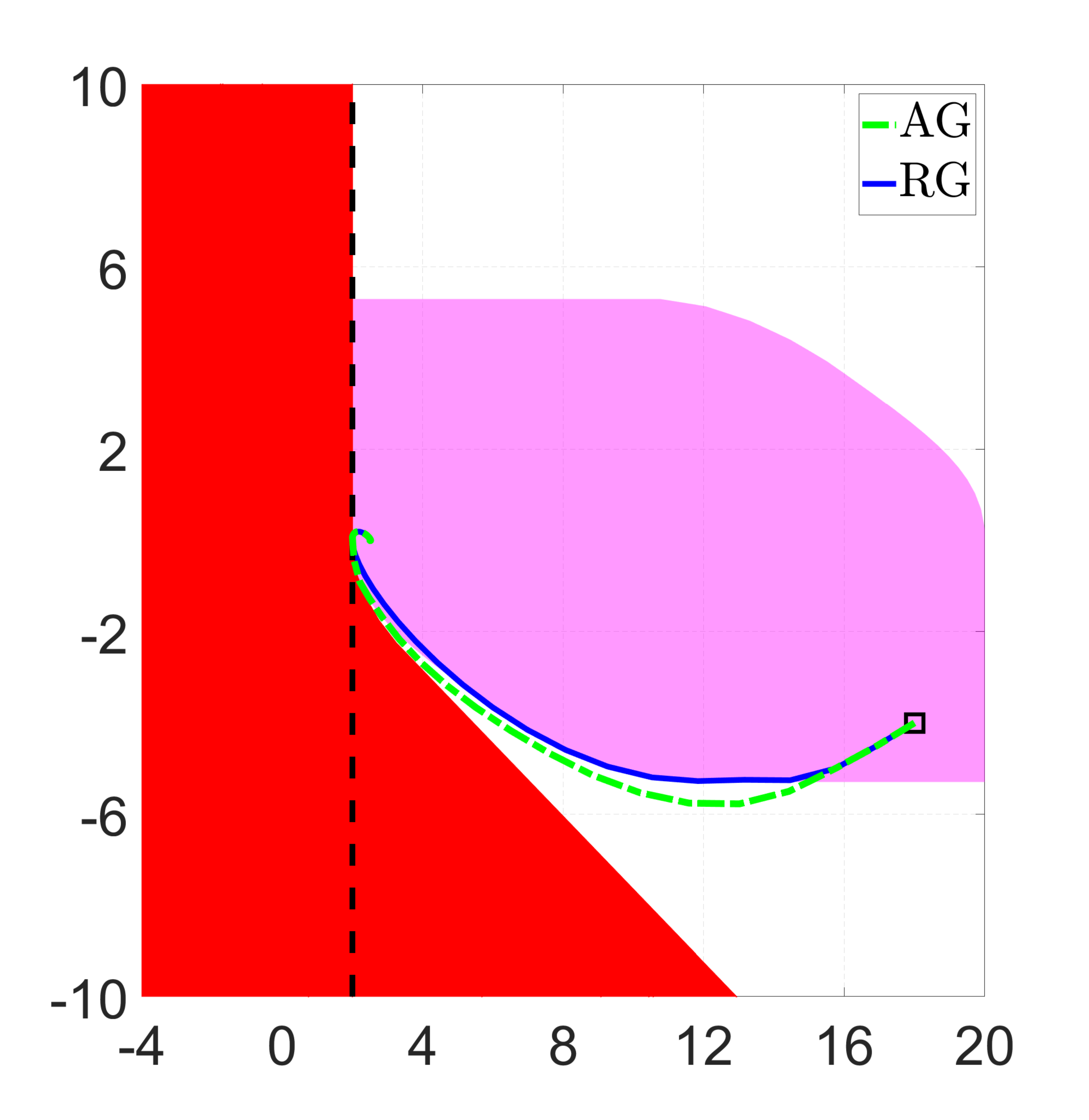,height=.26\textwidth}}  
\put(  120,  -10){\epsfig{file=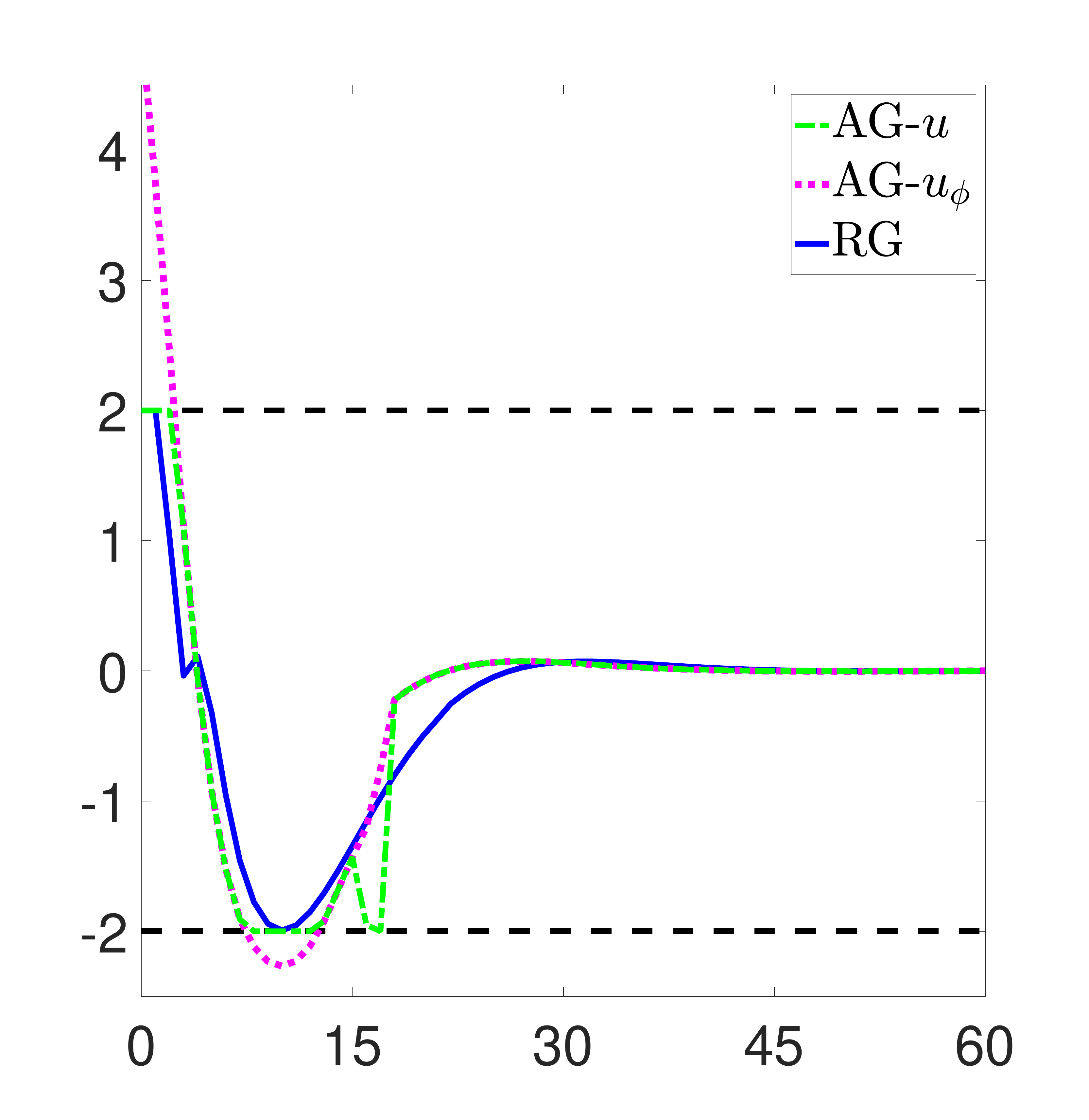,height=.26\textwidth}}  
\small
\put( 16, 116){(a)}
\put( 141, 116){(b)}
\put( 18, 64){$X_{\infty}$}
\put( 76, 64){$O_{\infty}$}
\put( 98, -10){$\Delta s$}
\put( -4, 100){$\Delta v$}
\put( 225, -10){$k$}
\put( 124, 100){$u$}
\normalsize

\end{picture}
\end{center}
      \caption{\small{Adaptive cruise control.}}
      \label{fig:Ex_1}
      \vspace{-0.07in}
\end{figure}
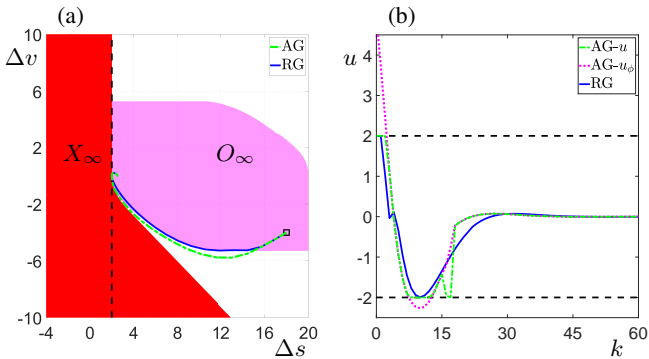

The simulation results are presented in Fig.~\ref{fig:Ex_1}. The state trajectory using AG (green dash-dotted) versus that using RG (blue solid) is shown in Fig.~\ref{fig:Ex_1}(a). It can be seen that the green curve converges to the desired steady state $(2.5,0)$ while being kept outside the set $X_{\infty}$ (red shaded) for all time. In particular, it sometimes rides on the boundary of $X_{\infty}$ but never enters into $X_{\infty}$. This guarantees the satisfaction of the state constraint $\Delta s(k) \ge 2$, whose boundary is marked by the black dashed vertical line. The blue curve also satisfies the constraint $\Delta s(k) \ge 2$ by being kept inside the set $O_{\infty}$ (magenta shaded) for all time and converges to the desired steady state. However, it is clear that the region where the state is allowed to reach using AG ($\mathbb{R}^2 \setminus X_{\infty}$) is strictly larger than that using RG ($O_{\infty}$), and this results in faster response and convergence with AG than RG, which can be seen from the control input trajectories plotted in Fig.~\ref{fig:Ex_1}(b). Also, both the control input trajectory using AG and that using RG satisfy the control constraints $-2 \le u(k) \le 2$. To achieve this, as well as to enforce the state constraint, AG sometimes modifies the nominal control $u_{\phi}$, the profile of which is shown by the magenta dotted curve.

\subsection{Omni-Directional Robot Obstacle Avoidance}

The dynamics of an omni-directional robot are modeled in continuous-time as 
\small
\begin{subequations}\label{equ:Omni_1}
\begin{align}
    \ddot{s}_1 &= u_1, \\
    \ddot{s}_2 &= u_2,
\end{align}
\end{subequations}
\normalsize
\!\!\! where $(s_1,s_2)$ represent the global positions of the robot in the $x$- and $y$-directions, and $(u_1,u_2)$ represent the accelerations and are the control inputs.

After being first written in first-order differential equations and then discretized with a sampling period $\Delta t = 1$ and assuming zero-order hold on the inputs, \eqref{equ:Omni_1} is converted to a discrete-time model in the form of \eqref{equ:system_1} with $(s_1,s_2,\dot{s}_1,\dot{s}_2)$ as the vector state and $(u_1,u_2)$ as the vector control input. To track a desired position $(s_{1,r},s_{2,r})$, a nominal control policy is defined as: for $i = 1,2$,
\small
\begin{equation}\label{equ:Omni_21}
    u_{i,\phi}(k) = \text{sat}_{R_i(k)} \left(K(i,:) \begin{bmatrix} s_1(k) - s_{1,r}(k) \\ s_2(k) - s_{2,r}(k) \\ \dot{s}_1(k) \\ \dot{s}_2(k) \end{bmatrix} \right),
\end{equation}
\normalsize
\!\!\! where $K$ is the LQR gain with $Q = \text{diag}(1,1,1,1)$ and $R = \text{diag}(1,1)$, $K(i,:)$ denotes its $i$th row, and $\text{sat}_{R_i(k)}(\cdot)$ is the saturation function to the range
\small
\begin{equation}\label{equ:Omni_22}
    R_i(k) \!=\! \left[\max\big(-2, -\frac{1}{\Delta t} (4 + \dot{s}_i(k))\big), \min\big(2,\frac{1}{\Delta t}(4 - \dot{s}_i(k)) \right].
\end{equation}
\normalsize
\!\!\! We remark that the control policy defined by \eqref{equ:Omni_21} and \eqref{equ:Omni_22} is a modified LQR control law equipped with the capability of enforcing the velocity and acceleration constraints
\small
\begin{equation}
-4 \le \dot{s}_i(k) \le 4, \quad -2 \le u_i(k) \le 2, \quad \forall\, k.    
\end{equation}
\normalsize

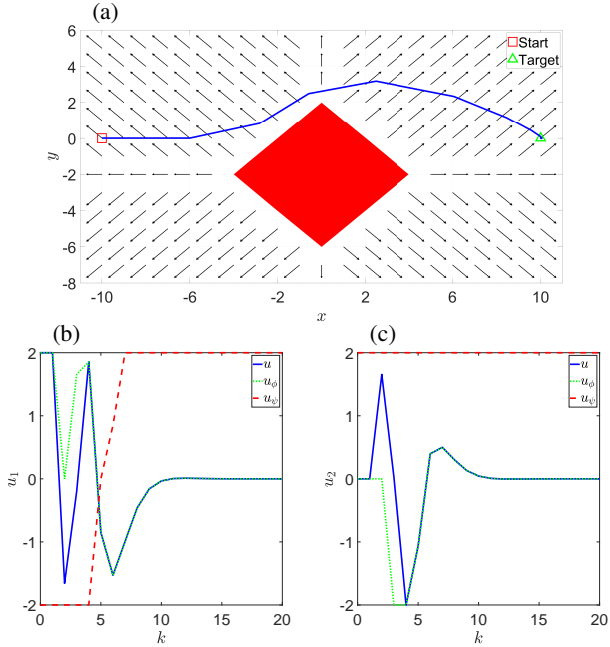
\begin{figure}[h!]
\begin{center}
\begin{picture}(240.0, 230.0)
\put(  0,  110){\epsfig{file=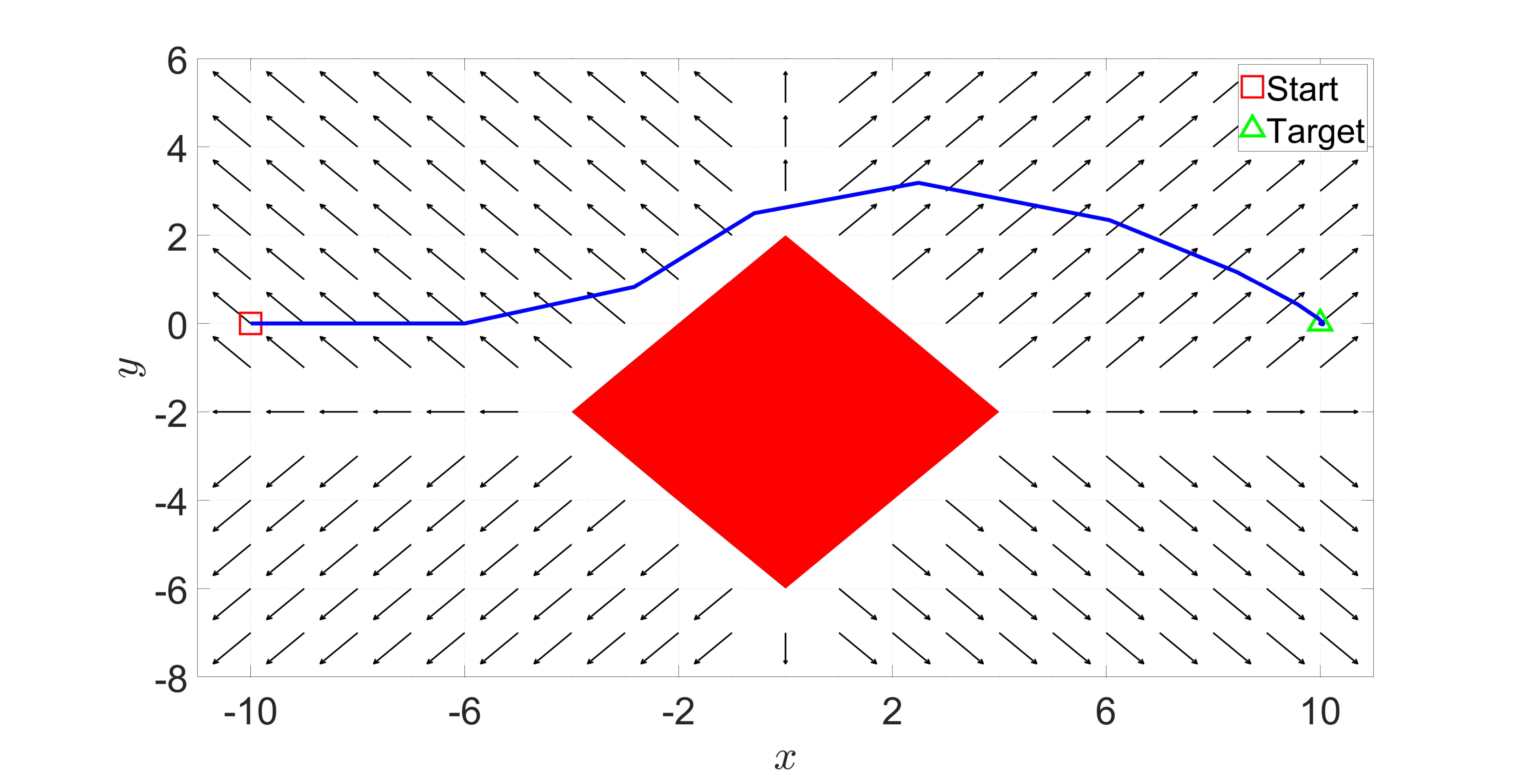,height=.24\textwidth}}  
\put(  0,  -12){\epsfig{file=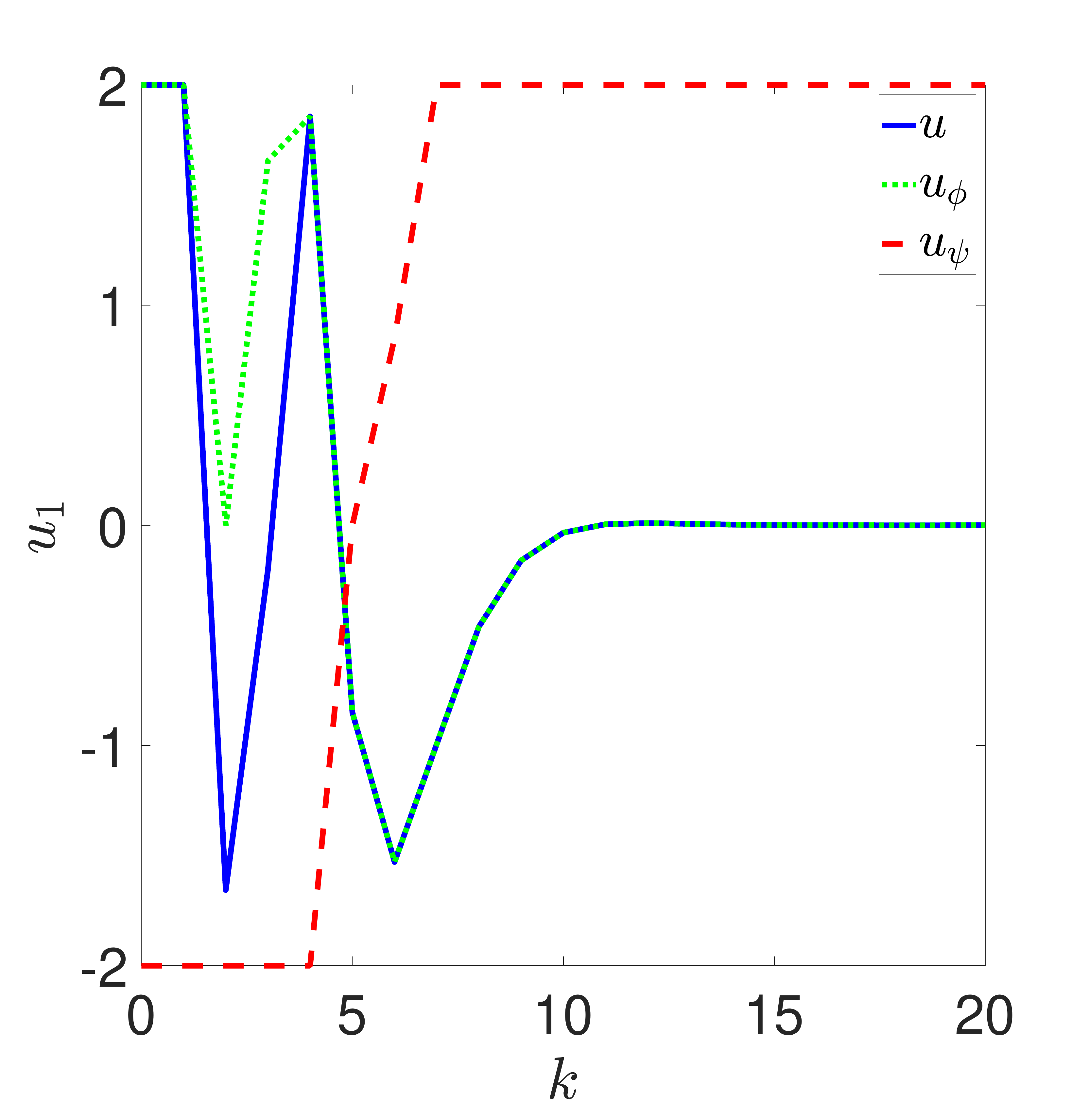,height=.24\textwidth}}  
\put(  120,  -12){\epsfig{file=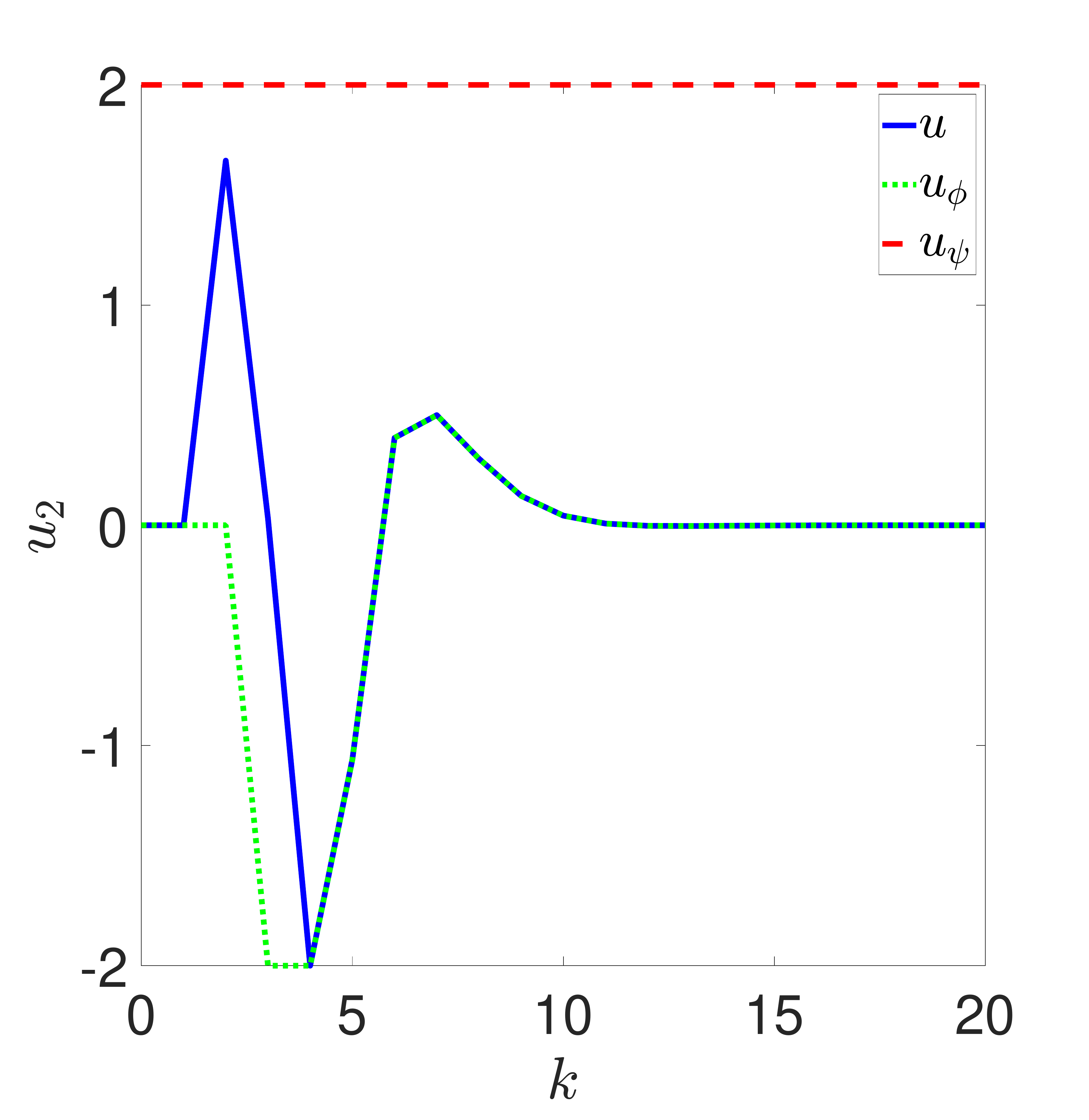,height=.24\textwidth}}  
\small
\put( 35, 226.5){(a)}
\put( 20, 104.5){(b)}
\put( 140, 104.5){(c)}
\normalsize

\end{picture}
\end{center}
      \caption{\small{Omni-directional robot obstacle avoidance.}}
      \label{fig:Ex_2}
      \vspace{-0.1in}
\end{figure}

We consider a scenario similar to the one studied in \cite{hermand2018constrained}. It is assumed that a diamond-shaped obstacle blocks the straight-line path from the robot's initial position $\big(s_1(0),s_2(0)\big) = (-10,0)$ to the target position $(s_{1,r},s_{2,r}) = (10,0)$, as shown in Fig.~\ref{fig:Ex_2}(a). To avoid collision with such an obstacle, we use an AG to supervise the control signal. In particular, the polytopic exclusion set $X_0$ is determined by the diamond-shaped obstacle and the velocity ranges $\dot{s}_i \in [-4,4]$, and the box-shaped control authority set $U$ is determined by the acceleration ranges $u_i \in [-2,2]$.

In this example, the online determination of the control inputs $(u_1,u_2)$ is through Algorithm~2, which relies on a safe-mode control policy $\psi$. We construct $\psi$ based on a simple repulsive-force field surrounding the obstacle \cite{latombe2012robot}, which is illustrated by the black arrows in Fig.~\ref{fig:Ex_2}(a). Specifically, the control $(u_{1,\psi},u_{2,\psi})$ is determined first as a vector along the direction of the arrow at the robot's current position with magnitude proportional to the arrow length (which is constant everywhere in the considered repulsive-force field), then saturated to the ranges $u_i \in [-2,2]$. We remark that one important difference of our approach from other obstacle avoidance approaches based on repulsive-force/potential fields is that the trade-off between position tracking and collision avoidance is optimized online through Algorithm~2 in our approach rather than pre-designed offline, which is typical in those approaches \cite{latombe2012robot}.

The simulation result is shown in Fig.~\ref{fig:Ex_2}. It can be seen from Fig.~\ref{fig:Ex_2}(a) that the robot safely travels from the start position (marked by the red square) to the target position (marked by the green triangle) without colliding with the obstacle. Figs.~\ref{fig:Ex_2}(b) and (c) show the control input histories. In particular, the green dotted curves correspond to the nominal policy $\phi$, the red dashed curves to the safe-mode policy $\psi$, and the blue solid curves are the optimized convex combinations of $\phi$ and $\psi$ obtained by Algorithm~2.

\section{Conclusions}\label{sec:5}

In this paper, we introduced the new add-on, supervisory scheme, {\it Action Governor} (AG), for discrete-time linear systems to satisfy exclusion-zone avoidance constraints. The AG enforces constraints by monitoring and modifying the nominal control signal based on set-theoretic techniques and online optimization. We established theoretical properties of the AG, discussed its computational realization, and used automotive and mobile robot related examples to illustrate its operation and effectiveness. Future research will address the cases where disturbances and model uncertainty are present, and extend the AG scheme to nonlinear systems.

\bibliography{ref}
\bibliographystyle{ieeetr}

\section*{Appendix}

\noindent {\em Lemma~1:} Given an increasing sequence of sets $X_k$ and an arbitrary $U$, we have $\big(\bigcup_{k=0}^{\infty} X_k\big) \sim U = \bigcup_{k=0}^{\infty} \big(X_k \sim U \big)$.

\noindent {\em Proof:} Note first that $Y_r = \big(\bigcup_{k=0}^{r} X_k\big) \sim U$ and $Z_r = \bigcup_{k=0}^{r} \big(X_k \sim U \big)$ are both increasing sequences of sets, and thus, their limits exist as $r \to \infty$ (in the set-theoretic sense). For each $r$, we have
\small
\begin{equation}\label{equ:lemma}
    Y_r = \bigcup_{j = 0}^{r} Y_j = \bigcup_{j = 0}^{r} \Big(\big(\bigcup_{k=0}^{j} X_k\big) \sim U\Big) = \bigcup_{j = 0}^{r} \big(X_j \sim U \big) = Z_r,
\end{equation}
\normalsize
\!\!\! where we have used the monotone increase of $X_k$ to derive the third equality of \eqref{equ:lemma}. Since $Y_r = Z_r$ for every $r$, it must hold that $\lim_{r \to \infty} Y_r = \lim_{r \to \infty} Z_r$, i.e., $\big(\bigcup_{k=0}^{\infty} X_k\big) \sim U = \bigcup_{k=0}^{\infty} \big(X_k \sim U \big)$. $\blacksquare$

\end{document}